\documentclass{nature3}

\usepackage{graphics}
\usepackage{graphicx}
\usepackage{amsmath}
\usepackage{amssymb}
\usepackage{color}

\title{Water dynamics: Relation between hydrogen bond bifurcations, molecular jumps, local density  \&
hydrophobicity
} 

\author{John Tatini Titantah$^{1}$ and Mikko Karttunen$^{2}$}

\date{\today}

\begin{document}

\maketitle

\begin{affiliations}
\item Department of Applied Mathematics, The University of Western Ontario, 
1151 Richmond St. North, London,  Ontario, Canada N6A~5B7
\item Department of Chemistry and Waterloo Institute for Nanotechnology, 
University of Waterloo, 200 University Avenue West, Waterloo, Ontario, Canada N2L~3G1 

\end{affiliations}

\baselineskip24pt
\maketitle 

\begin{abstract}
Structure and dynamics of water remain a challenge. Resolving the properties
of hydrogen bonding lies at the heart of this puzzle. Here we employ 
\textit{ab initio} Molecular Dynamics (AIMD) simulations over a wide
temperature range. The total simulation time was $\approx 2$\,ns. 
Both bulk water and water in the presence of a small hydrophobic
molecule  were simulated. We show that large-angle jumps and bond bifurcations 
are fundamental properties of water dynamics and that they are intimately coupled to both local 
density and hydrogen bond stretch oscillations in scales from about 60 
to a few hundred femtoseconds: Local density differences are the driving force for bond bifurcations and the
consequent large-angle jumps. The jumps are intimately connected
to the recently predicted energy asymmetry. Our analysis also appears to confirm the existence of 
the so-called negativity track provided by the lone pairs of electrons on the oxygen atom 
to enable water rotation.  
\end{abstract}

\section{Introduction}

Static and dynamic properties of both bulk and solvation water have garnered 
a lot of attention including several conflicting hypotheses. 
For example, the picture that water rotation can be desribed by simple Debye rotational
diffusion was shown to be too simplistic by Laage and Hynes who demonstrated, 
using computer simulations, that instead of smooth rotation, water molecules 
undergo large-angle jumps\cite{laagescience}. This has since been confirmed and
by experiments\cite{ji10} and independent simulations\cite{titantah2012}.
In a recent contribution, \textit{ab initio} simulations of K\"uhne 
and Khaliullin\cite{kuhne13} resolved the controversy regarding
the interpretation of a series of x-ray  experiments\cite{wernet04,tokushima08} 
that suggested water having anisotropic  structure. Their simulations showed that 
the structure of water is tetrahedral but the energetics of the hydrogen bonds (HBs) 
are asymmetric. 

The dynamics and structure of hydrogen bonding lie at the heart of 
the above and other anomalies of water, and, importantly, water's properties 
as a solvent in biological systems. They give rise to 
the collective nature of hydrophobic hydration which, for example, allows proteins to explore 
their configurational space; direct point-to-point binding would prevent that. Thus,
understanding how water interacts with other molecules, in particular with 
hydrophobic groups, is vital for understanding processes in molecular 
biology\cite{Pal2004,Chaplin:06aa}.  
Hydrophobic hydration has been intensely studied and debated, but
the femtosecond mid-infrared (fs-IR) experiments of Rezus and Bakker\cite{rezus07}, 
that showed that OH rotational motion is reduced within the first hydration shell, 
revived the discussion by showing that solvation water around small
hydrophobic groups behaves distinctly differently from bulk water and
having characteristic described by the so-called Iceberg model of hydration\cite{Frank1945}.

We have performed AIMD study of 
the the behaviour of both bulk and solvation water. The latter was done in
the presence of tetramethylurea (TMU). Dispersion corrections\cite{grimme04,hujo2013}
were applied to avoid the well-known overstructuring problem. The total simulation time was
$\approx 2$\,ns and each individual simulation at least  100\,ps after equilibration. 
We show the connection between the energetic asymmetry in hydrogen bonding and the 
large-angle jumps water molecules execute. We show that the time scale
for reaching equilibrium, and hence the molecular jumps, depend sensitively 
on the local density.

\section{Results}
\subsection{Frequency shift.}

We first verified the results of 
K\"uhne and Khaliullin 
for the  anisotropy in the hydrogen bond strength\cite{kuhne13}. 
The  asymmetry and distortion they observed in hydrogen bond energetics can also 
be detected through  the OH stretch vibrations
of water.  A time series analysis was performed to determine the 
time-dependent OH stretch vibrational frequency of neat H$_2$O and H$_2$O 
hydrating the TMU molecule. We adopted the method used by 
Mallik \textit{et al.}\cite{mallik08} and performed a  continuous wavelet 
transform of the trajectories. This method allows for
spectral analysis of systems whose frequencies vary strongly with time.
 
\begin{figure}
\begin{center} 
\includegraphics[viewport=170 60 450 400,width=10cm]
{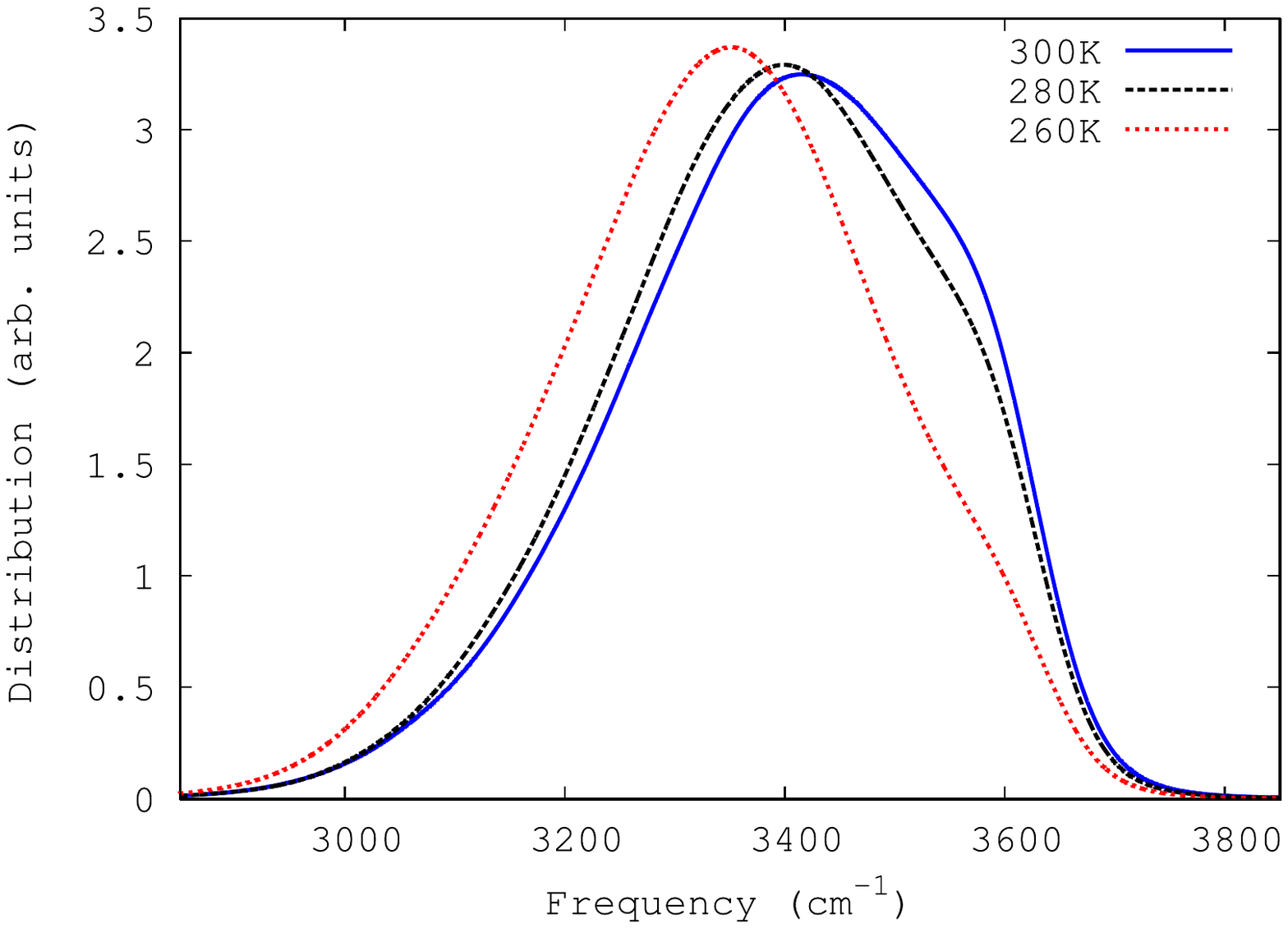}
\end{center}
\caption{The OH-stretch frequency distribution obtained using 
the wavelet transform approach of Mallik \textit{et al.}\cite{mallik08} 
for bulk water at temperatures of 260\,K, 280\,K and 300\,K. 
Because of the 6\%~underestimation of the line centre at 300\,K 
(with respect to the experimental value\cite{loparo04} of about 3400 cm$^{-1}$), 
the frequency axis has been rescaled by a factor of 1.06.\cite{loparo04}
}
\label{fig:freq-distr}
\end{figure}

For bulk water, the line-width -- the fingerprint of 
the dynamics -- was found to be 350 cm$^{-1}$. This  is in 
excellent agreement with experiments:
IR measurements and Raman spectroscopy give 350 cm$^{-1}$ and 350-400 cm$^{-1}$, respectively\cite{pershin04}  
and fs-IR studies of Fecko \textit{et al.}\cite{fecko05a} obtained 350 cm$^{-1}$.
Classical water models, such as TIP4P 
and the fluctuating charge TIP4P-FQ, fail to account for the 
experimental width\cite{harder05}.  The OH stretch-frequency 
distributions obtained in this work for bulk water at supercool, low and room 
temperature are shown in Fig.~\ref{fig:freq-distr} 
demonstrating an asymmetry towards lower frequencies. 
This is in good agreement with depolarized Raman spectroscopy measurement on liquid water\cite{skinner09}.

\begin{figure}
\begin{center} 
\includegraphics[viewport=160 80 720 500, width=14cm,angle=0,clip=false]
{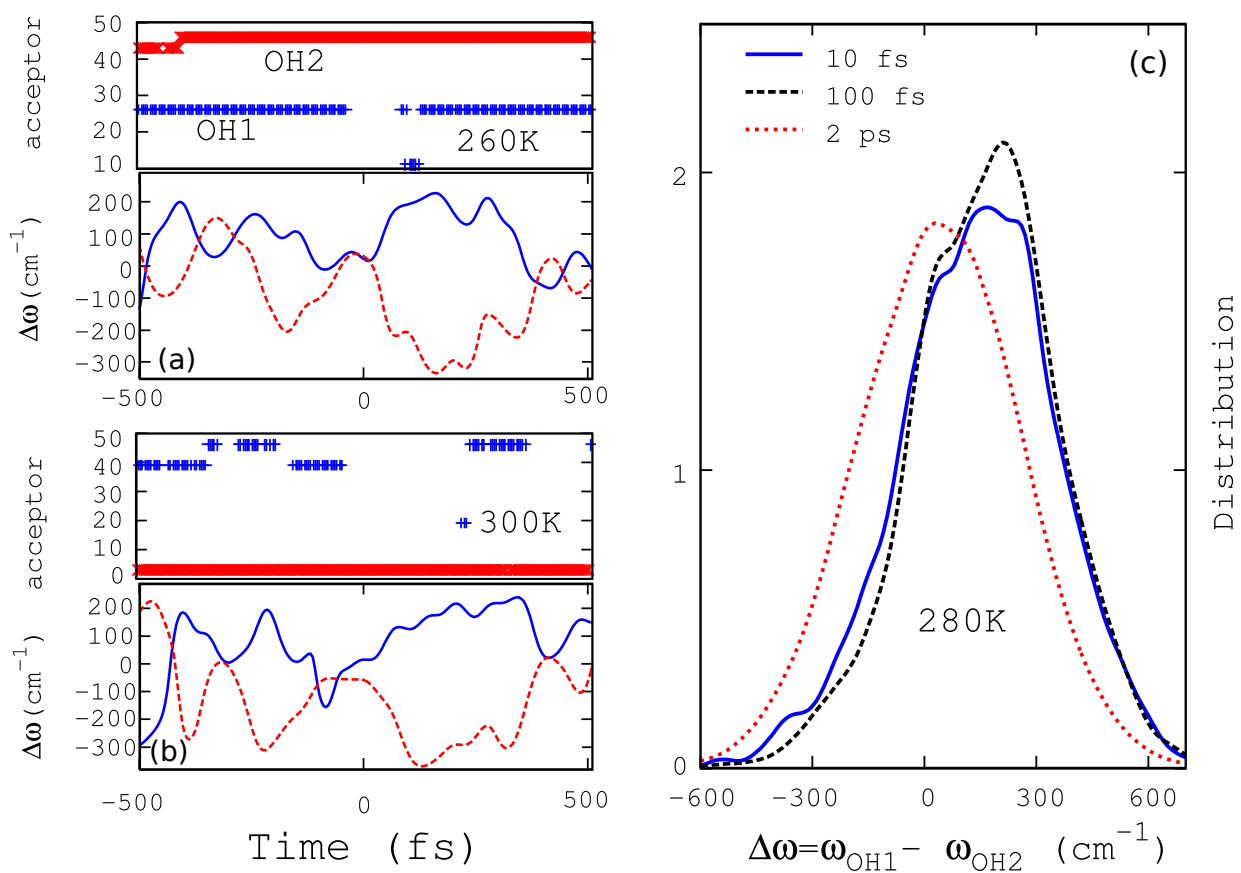}
\end{center}
\caption{
The upper parts of (a) and (b) give the identities representing the instantaneous hydrogen-bond-accepting 
water molecules (numbering from 1 to 54). OH1 accepts an HB 
at $t=0$. OH2  stands for the other OH group. 
The lower parts of panels (a) and (b) show the corresponding OH frequency shifts. 
The full lines correspond to the the OH group that lost a hydrogen bond at time 
$t=0$ (jump instant) and the dashed line is for the other OH group (OH2) that remains hydrogen-bonded. 
(c) The distribution of the difference between the shifts $\Delta \omega = \omega_{OH1}-{\omega_{OH2}}$ 
at various times from the instant of the HB loss on OH1. Results are for bulk water.}
\label{fig:shift-acceptors}
\end{figure}

Next, we examined the connection between the anisotropy in HB strength
and the large-angle jumps water molecules execute\cite{laagescience,ji10,titantah2012}.
We considered water molecules that donate two HBs to two other water 
molecules through their OH groups. Those OH groups are labelled as OH1 and OH2. 
We monitored their OH stretch frequencies before 
and after one of them loses its HB partner or performs 
a large-angle jump to bind with a new partner. This analysis 
is demonstrated for bulk water in Figs.~\ref{fig:shift-acceptors}a,b 
for two temperatures, 260\,K and 300\,K. 

Figures~\ref{fig:shift-acceptors}a,b show the frequency shifts with respect to the line center.
The  upper panels show the identities of the water 
molecules  (numbering from 1 to 54) that accept HBs from OH1  and OH2. 
The commonly used structural definition\cite{titantah2012} 
based on O-O distances ($d_\mathrm{O-O}<3.5$~\AA) and OHO angle ($\theta\le$30$^\circ$)
was used to determine HBs.
The choice of $d_\mathrm{O-O}$ corresponds to the first minimum of the O-O radial distribution function.
The  large-angle jump takes place at time $t=0$. 
The  instant of the jump is defined as the time when the distance between 
the oxygen atom of the water molecule donating the HB and the oxygen atom of the 
HB-accepting water before the jump, and that between the former and 
the  oxygen atom of the new HB partner are equal\cite{laagescience,titantah2012}.  
Figure~\ref{fig:shift-acceptors}a shows the situation 
at 260\,K: OH1 is hydrogen bonded to water molecule number 27 before $t=0$ 
while OH2 is bonded to water number 46 (upper panel of Fig.~\ref{fig:shift-acceptors}a). 
At $t=0$  OH1 loses its HB with 
water-27. A dangling bond remains for 150\,fs at which time the 
HB with water-27 is re-formed and an unstable (bifurcated)  HB with water-11 emerges. 
Water-11 eventually breaks away while OH1 maintains its HB with water-27.  
During this time, OH2 remains H-bonded to water-46. As expected, the OH  group 
that loses the HB gains vibrational energy and its stretch frequency 
blue-shifts (lower panel of Fig.~\ref{fig:shift-acceptors}a).  Simultaneously, the OH group still hydrogen 
bonded shifts further into the red. Vibrational energy is 
transferred from the latter to the former.  This behaviour is seen 
for all such water molecules and at all temperatures, Fig.~\ref{fig:shift-acceptors}b 
demonstrates this at 300\,K.
This confirms the asymmetry reported by K\"uhne and Khaliullin and, importantly, 
shows for the first  time its relation to the dynamics of water and the inherent larger-angle jumps all water molecules execute.

The time evolution of this frequency shift is better monitored by 
performing time averages for the time spanning from the instant of an HB 
loss. At each time we calculate the difference 
between the frequencies of OH1 and OH2, $\Delta \omega_\mathrm{OH}$. The distributions of this 
shift for times of 10\,fs, 100\,fs and 2\,ps at T=280\,K are shown 
in Fig.~\ref{fig:shift-acceptors}c.
At shorter times (10 and 100\,fs)
the distribution is asymmetric and dominantly positive 
(reminiscent of the large-angle jump that took place at $t=0$).
At longer times (picosecond range) it approaches a symmetric shape.

\subsection{Temporal behavior of the frequency shift.}
\begin{figure}
\begin{center} 
\includegraphics[viewport=100 60 760 500, width=16cm,angle=0,clip=false]
{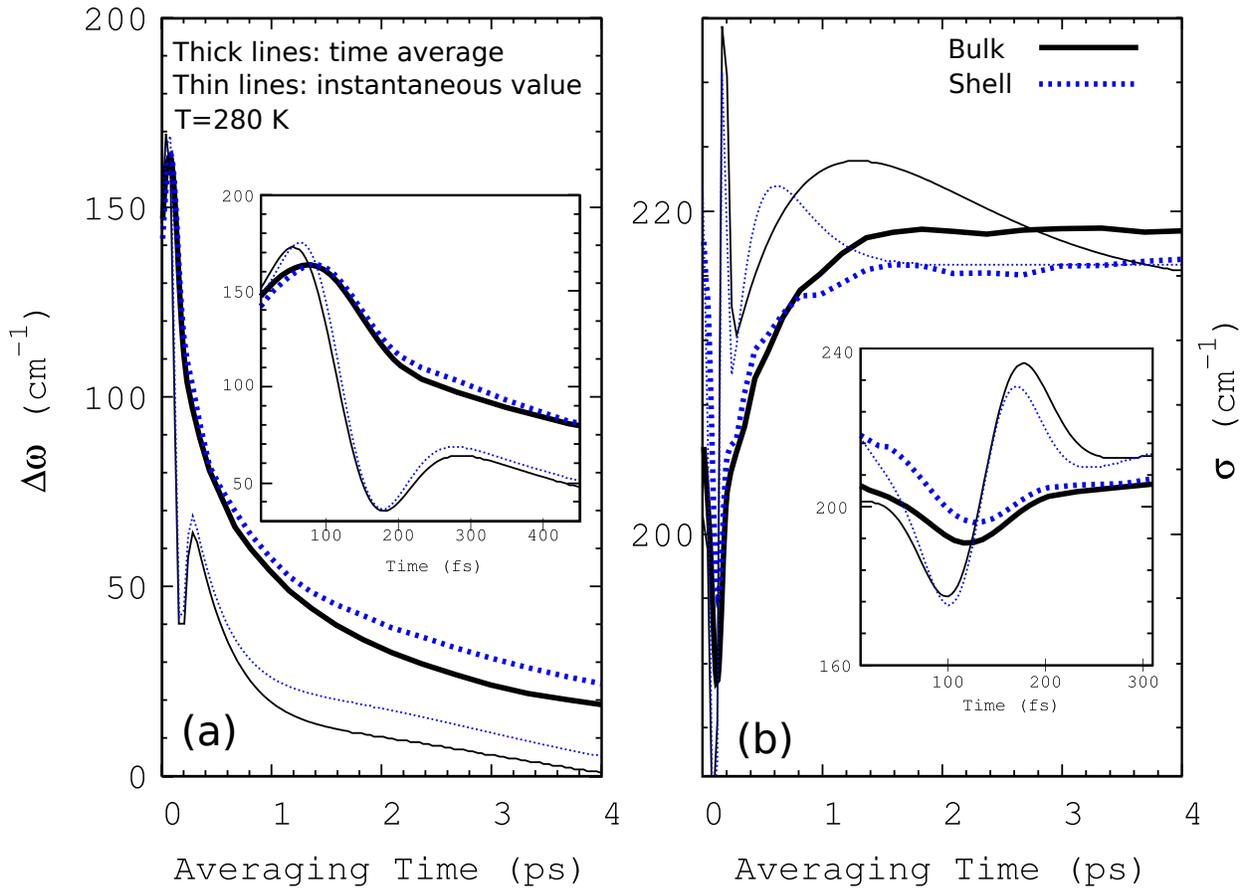}
\end{center}
\caption{(a) Time-averages of the frequency difference $\Delta \omega = \omega_{OH1}-{\omega_{OH2}}$  
for times from the instant of an HB loss by OH1. (b) Time-evolution of the standard deviation 
of the the distribution $\sigma=\sqrt{\left<\left(\omega_{OH1}-\omega_{OH2}\right)^2\right>}$ 
of the frequency difference.  Results are shown for  T=280\,K. 
Insets: The short time behaviours of $\omega_{OH1}-\omega_{OH2}$ and $\sigma$. 
Full lines: Bulk water.  Dashed lines: Water solvating the TMU molecule. 
Thick lines: Time averages whereas. Thin lines: The corresponding instantaneous values.}
\label{fig:sigma}
\end{figure}

To better characterize the temporal behaviour, 
we studied the time-dependence of the position of 
the line-center (Fig.~\ref{fig:sigma}a) and the standard deviation  (Fig.~\ref{fig:sigma}b)
of this distribution.
Figure~\ref{fig:sigma}a
shows the time averaged shifts as thick lines for bulk 
water (full lines) and solvation shell water around the TMU molecule (dashed lines).  
The shift increases initially, passing through a maximum 
at $80\pm5$\,fs and then decreases rapidly 
before entering a slower regime beyond 200\,fs. 
The long time tail follows a 
bi-exponential decay for both bulk and solvation water
with the shorter time ranging between 200 and 700\,fs 
for temperatures from 350\,K to 260\,K. At 300\,K this time was found to be 300$\pm40\,$\,fs  
for bulk water and is in excellent accord with 
intramolecular energy transfer rate of 300$\pm60\,$ fs for transfer between the free O-D and 
the hydrogen-bonded O-D group of the same molecule as found using 2D surface vibrational 
spectroscopy study on water/air interface\cite{zhangnature2011}.

The instantaneous shifts $\Delta\omega(t)$ and widths $\sigma(t)$ can be obtained 
from the time averages.
This was done on smooth functional fits through the time averages. The corresponding instantaneous values are 
shown as thin lines in Figs.~\ref{fig:sigma}a~and~b.  
They show short-time dynamics that is different from the time averages: 
The frequency shift in Fig.~\ref{fig:sigma}a increases initially up to 60\,fs 
where it passes through a maximum and decreases to a minimum 
at  180$\pm$10\,fs. It then increases and passes through a broader local maximum at about 270$\pm$\,20\,fs.
The line-width in 
Fig.~\ref{fig:sigma}b also reveals
an oscillation with a period of $\approx 180$\,fs.

The time of 80\,fs deduced from the time averages (Fig.~\ref{fig:sigma}a), is practically the same as that reported by K\"uhne ~\textit{et al.}  
for the oscillatory anisotropy of the HB strength. The time of 60\,fs apparent in instantaneous values (inset in Fig.~\ref{fig:sigma}a)
is  the time between  the instant  of the bifurcated H-bonded state 
and the instant when one of the HB-accepting water molecules 
breaks loose from the bifurcated state. 
This is the same as the time of the loss of frequency correlation which 
has also been demonstrated using fs-IR spectroscopy to be 50\,fs\cite{kraemer08}.
We will return to this issue later.

\begin{figure}
\begin{center} 
\includegraphics[viewport=100 60 760 500,width=16cm,angle=0,clip=false]
{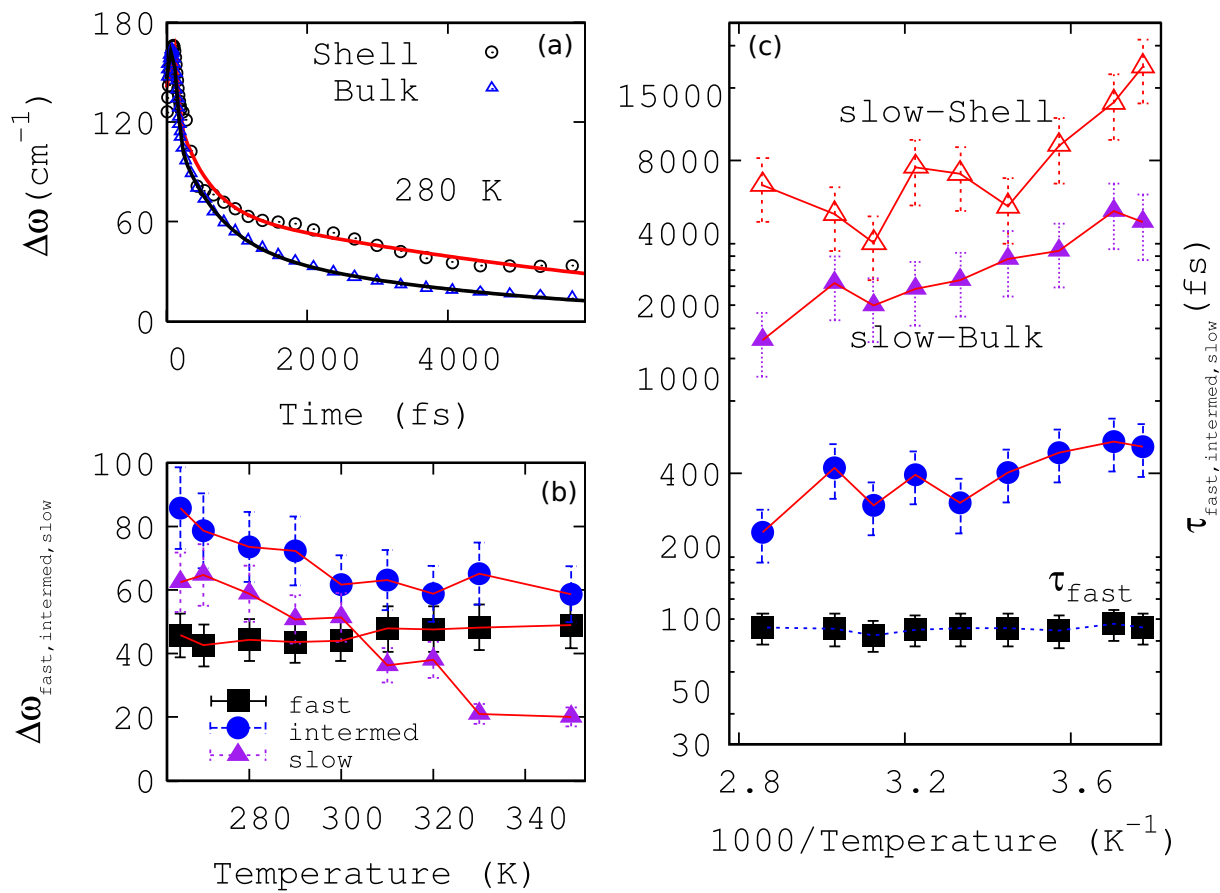}
\end{center}
\caption{(a) Time dependence of the frequency shift $\Delta\omega$ (in cm$^{-1}$) for bulk water (triangle) and solvation water (circle) at 280\,K. They are fitted as sum of a Gaussian centred at $\tau_{fast}=\tau_0$ with standard deviation $\sigma_0$ and two exponential tails with time constants $\tau_{intermed}$ and $\tau_{slow}$. These are shown as the dotted lines for bulk water. (b) Temperature dependence of the magnitude of each of the three components in (a) for bulk water. (c) Inverse-temperature effect on the relaxation times described in (a). The fast and intermediate times are given for bulk water and the slow times are given for both bulk and solvation water.}
\label{fig:shift-fits}
\end{figure}

A fit of the time-dependence of the frequency shift 
for bulk water at 280\,K is shown together with the shift for 
solvation shell water  in Fig.~\ref{fig:shift-fits}a.  
A  combination of a Gaussian centred at $t=\tau_0$ with 
standard deviation $\sigma_0$  and two exponential terms 
(corresponding to the intermediate and slow terms) was used 
to model the time-dependence.  The magnitudes of the shifts 
corresponding to the fast, intermediate and  slow behaviours are shown in 
Fig.~\ref{fig:shift-fits}b for various temperatures. 

The fast component 
of magnitude 45$\pm$5 cm$^{-1}$ (see ~Fig.~\ref{fig:shift-fits}b), was found to be virtually 
independent of temperature. 
In contrast, the intermediate and the slow components are 
strongly temperature dependent. While the former is a consequence of 
temperature induced rearrangement of the destabilized local water structure 
that  results from the large-angle HB exchange events, the latter may be a 
diffusion-related energy transfer or spectral diffusion as the timescale involved is 
of the order of diffusional times (picoseconds). We also 
found that the diffusion coefficient of the corresponding water 
molecules follow a similar temperature behaviour.  
The shift corresponding to the intermediate rate dominates over the other two (Fig.~\ref{fig:shift-fits}b), 
indicating that this channel of vibrational energy transfer is the most prevalent. 
Its timescale that spans from 200 to 700\,fs is also comparable 
to vibrational energy transfer times in water\cite{yang11}.
Figure~\ref{fig:shift-fits}c shows the temperature dependence 
of the three timescales involved in the dynamics of the frequency shift. 
We also performed the above analysis for water molecules solvating the 
TMU molecule and found that the magnitude and timescale of the fast and 
intermediate energy transfers events are very similar to those of bulk water. 
The slow component behaves differently: As Fig.~\ref{fig:shift-fits}c shows, the slow time is seen to increase by a factor 
of 2-4 for solvation shell water around hydrophobic TMU 
(compare the upper two curves of Fig.~\ref{fig:shift-fits}c), i.e., at low temperatures the hydrophobic 
effects slows down the intramolecular energy transfer. 
An Arrhenius fit using the slow time for bulk water yielded an activation energy of 13$\pm$1\,kJ/mol. 
This compares very well with the value of 12 kJ/mol reported as the anisotropy barrier between the strong and weak  HB donors\cite{kuhne13}.

\subsection{Creation and relaxation of energy anisotropy.}
\begin{figure}
\begin{center} 
\includegraphics[viewport=100 60 760 500,width=16cm,angle=0,clip=false]
{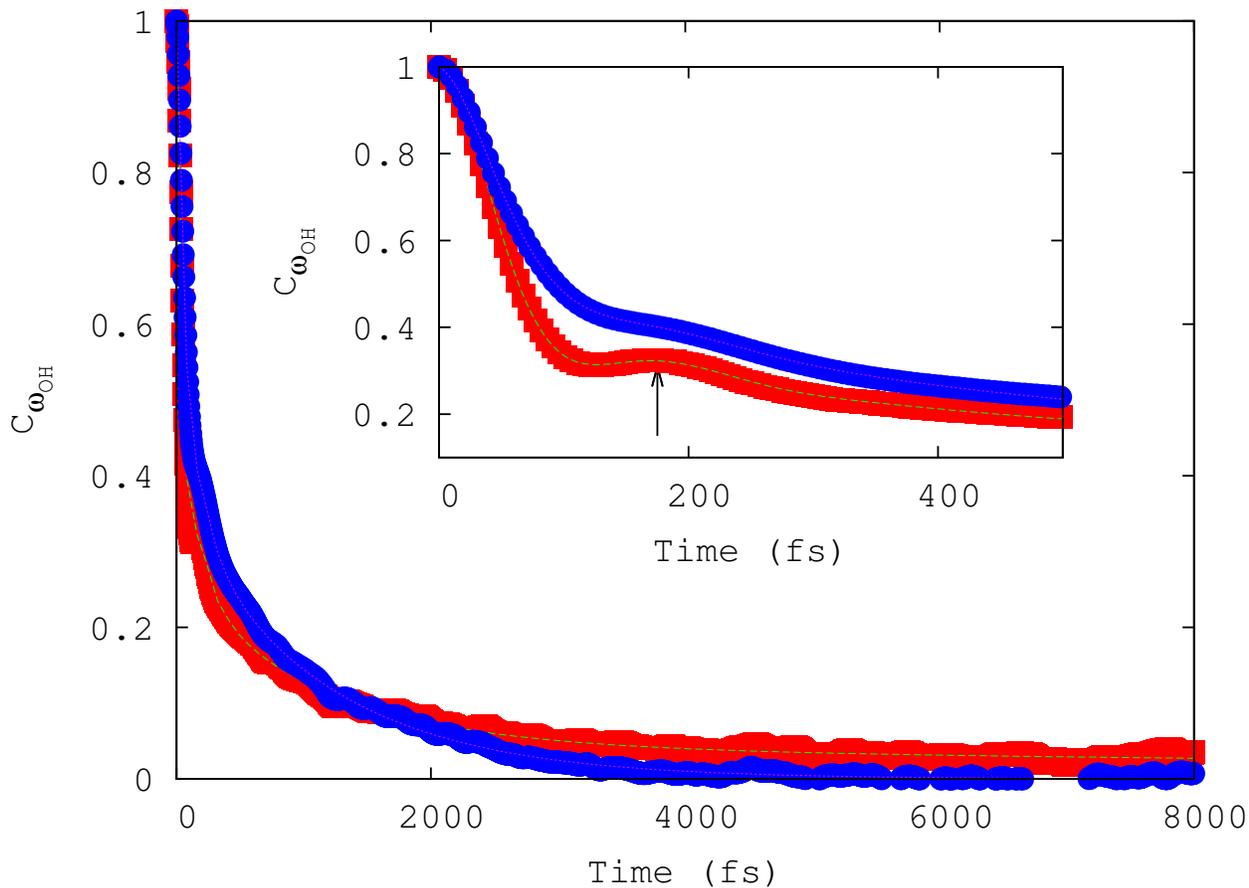}
\end{center}
\caption{OH stretch frequency-frequency time correlation of OH groups initiated in H-bonded state (red) and those in non-bonded state (blue). The inset shows the short-time behaviour revealing a bump at 175\,fs especially for the H-bonded OH groups. The short time behaviour shows a rapid loss of frequency correlation with a timescale of 60$\pm$5\,fs. Notice that while the blue end is slow at short times, the red end is slow at longer times.}
\label{fig:oh-corr}
\end{figure}

As discussed above, the data shows an oscillatory behaviour for HB strength with a characteristic period of 180 $\pm$10\,fs. This 
is in excellent agreement with HB  strength oscillatory dynamics 
found in fs-IR spectroscopy with a period of 170\,fs\cite{fecko03}.  This oscillation period 
can also be seen from the frequency-frequency time-correlation function  
performed in two frequency regimes: One for OH groups initiating from an H-bonded state (red-end of the spectral line) 
and another starting in the non-bonded or weakly bonded state (blue-end). The bonded OH groups are 
those whose stretch frequencies are below the line-centre (3400 cm$^{-1}$) and the non- or weakly-bonded ones are those 
with higher frequencies. This definition is approximate, as those groups whose frequencies are around 
the line-centre may show both bonded and non-bonded character. The correlation functions for bulk 
water at 300\,K are shown in Fig.~\ref{fig:oh-corr}. That associated with the relaxation of 
the H-bonded groups clearly shows  a strongly damped oscillatory behaviour as the bump at 175\,fs 
indicates with the 
damping factor of 60$\pm$5\,fs (through fitting with a combination of two exponentials and a damped oscillatory function) 
being equal to the life-time of the fast  
component of the instantaneous intramolecular frequency shift elaborated above and shown in the inset of Fig.~\ref{fig:sigma}a.
This damping factor of 60$\pm$5\,fs compares well with 2D-IR spectroscopy study 
of  water where a value of 50\,fs  has been reported for the  initial loss of frequency correlation\cite{kraemer08}.
Notice that while the blue end is slower than the red end at short times (Fig.~\ref{fig:oh-corr}), the red end becomes slower 
at longer times. This is in agreement with  experimental data\cite{rezus06,cringus05,kraemer08}.

K\"uhne ~\textit{et al.} pointed out that the relaxation of the asymmetry is influenced 
by the low frequency librational motion of the water molecule. They also stated that
"some second strongest interactions are weakened by distortions to such an extent 
that back donation to (from) a nearby donor (acceptor) becomes the second strongest 
interaction."\cite{kuhne13}  As our results and the above discussion shows,
this is fully consistent with the large-angle jump mechanism. 
We argue further that the connection of
the energy asymmetry is even more fundamental: Our results show that the large-angle 
rotational jumps performed by water molecules during HB exchange play a major role in the creation 
and the relaxation of this anisotropy. The dynamics  of the two 
OH groups of each water molecule are strongly coupled. The effect of local density on this coupling will 
be explored next.

\subsection{Local density differences give rise to jumps.}
We now study the connection between  
the fundamental process of 
vibrational energy transfer and the large-angle jumps.
In particular, we will show that an increase  in the local density around the central rotating water molecule 
drives the large-angle jumps.  For each water molecule, the local water density is defined as the mass-density  in a 
sphere of radius 3.5~\AA~around the oxygen atom of the water molecule of interest (both bulk and solvation shell). As mentioned 
earlier, the instant of the HB jump is defined as the time when the two HB-accepting water molecules 
forming a bifurcated HB with OH1 (denoted as O$^a$ and O$^b$) are equidistant from the oxygen atom of 
the HB-donating molecule (denoted O$^*$). O$^c$ denotes the water molecule accepting an HB 
from the other OH group of O$^*$ (OH2). 

\begin{figure}
\begin{center} 
\includegraphics[viewport=100 60 760 500,width=16cm,angle=0,clip=false]
{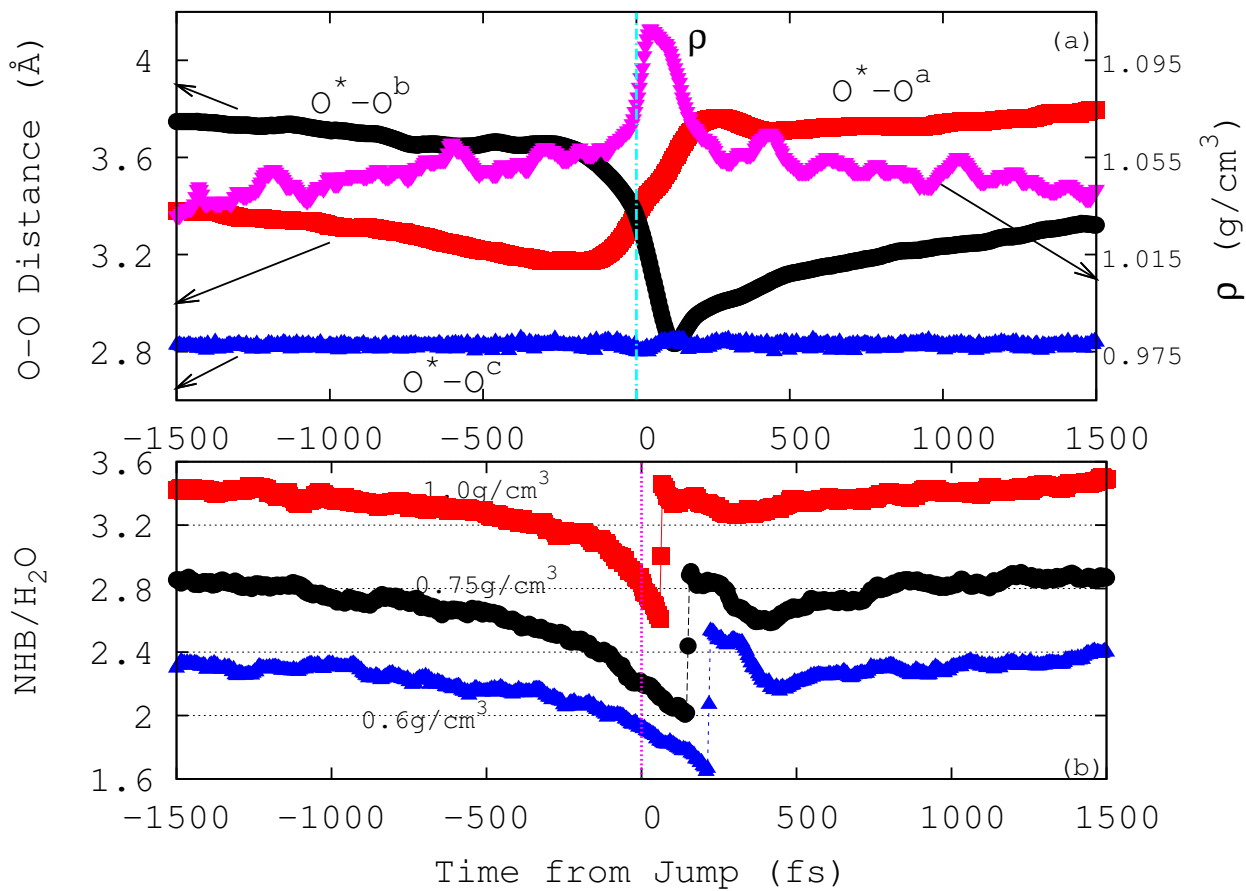}
\end{center}
\caption{(a) The time evolution of the local density around the central rotating water O$^*$ around the instant of the HB jump of its OH group (OH1) and the distance between O$^*$ and the HB-accepting water before the jump O$^a$, that after the jump O$^b$  and the water accepting accepting HB from the other OH group (OH2) of the central rotating water (T=280\,K). (b) Density effect on the number of  HB per water around the HB jump. The vertical lines denote the instant of the jump.}
\label{fig:o-o-dist}
\end{figure}

Figure~\ref{fig:o-o-dist}a shows the mass density around an HB-donating water molecule O$^*$ and
the distances O$^*$-O$^a$, O$^*$-O$^b$ and O$^*$-O$^c$.  These are averages over all OH1 groups and 
all jumps during the entire simulation run. The time $t=0$ is the instant of the jump. 
The left axis corresponds to the O-O distances and the right 
to local water density $\rho(t)$ (see the arrows in Fig.~\ref{fig:o-o-dist}a). While O$^*$-O$^a$ and 
O$^*$-O$^b$ distances undergo rapid changes,  the distance O$^*$-O$^c$ stays at the same value (of about 2.8~{\AA}). 
This shows  that the jump does not perturb the OH2 group which remains hydrogen bonded.  
The fact that large-angle jumps\cite{ji10,laagescience,titantah2012}   of 40-50$^\circ$ 
occur while one of the donated HBs and two of the accepted HBs are intact 
(Fig.~\ref{fig:o-o-dist}b shows that just before the jump the water molecule concerned has roughly 3 HBs), 
appears to confirm the existence of a negativity track\cite{noam12} provided by 
the lone pairs of electrons on the oxygen atom of the rotating water molecule. 
This permits a gliding motion of the HB-donating protons of neighbouring water molecules.

\begin{figure}
\begin{center} 
\includegraphics[viewport=100 60 760 500,width=16cm,angle=0,clip=false]
{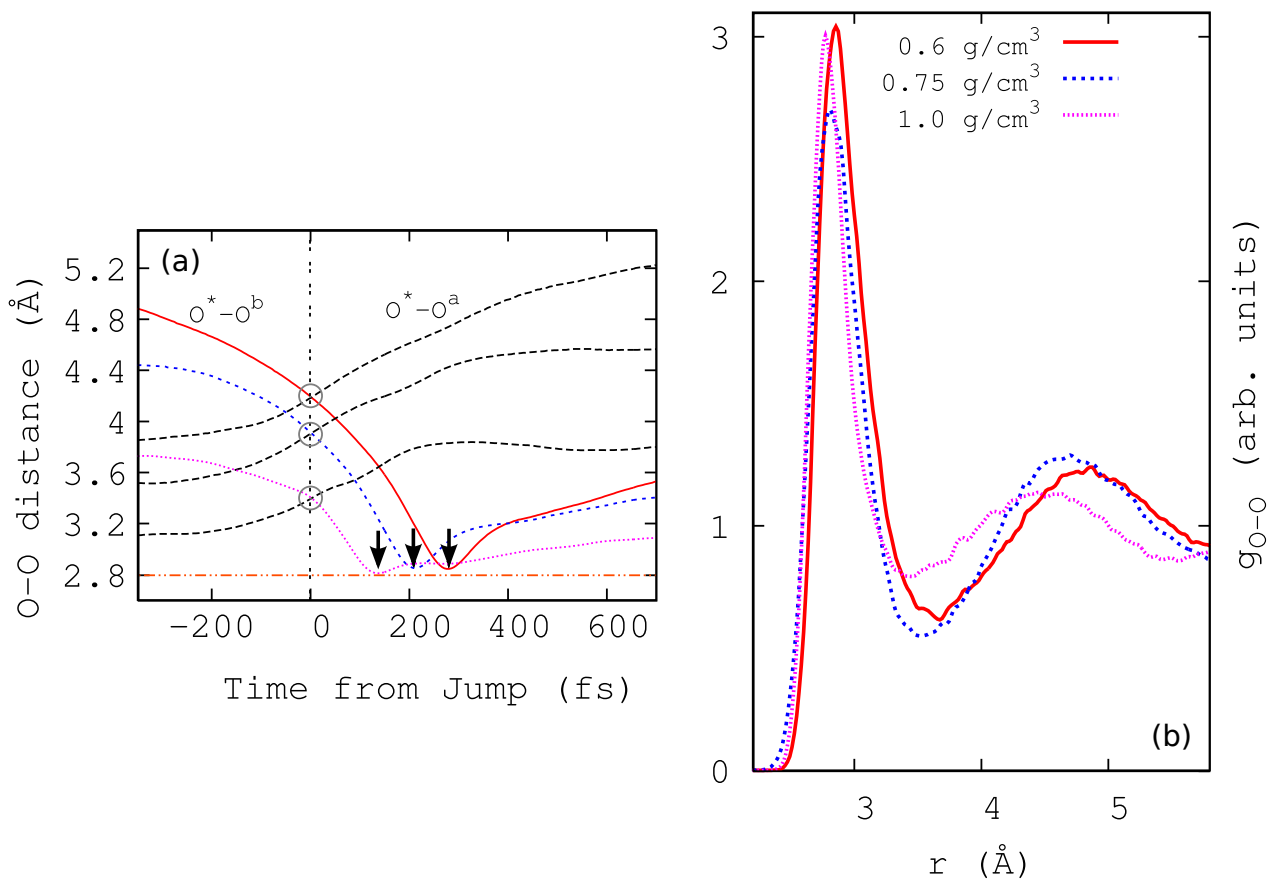}
\end{center}
\caption{
(a) The time evolution of  the distance between O$^*$ and the HB-accepting water before the jump O$^a$ (dashed lines) 
and  that after the jump O$^b$ (coloured lines) at 300\,K at different densities (the colour code is the same as in (b)). 
Notice that the bifurcated state occurs at  O-O distances that increase as density decreases 
(open circles on the vertical line) causing the newly form O-O pair to reach equilibrium distance 
of 2.8~{\AA} at longer times as indicated by the arrows. 
(b) the O-O pair correlation that shows first neighbor shell at almost the same value of 2.8 {\AA} 
and second neighbor at 4.5$\pm$0.2~{\AA}.  The vertical lines in (a) and (b) denote the instant of the jump.}
\label{fig:o-o-dist-gr}
\end{figure}

Immediately after the jump,  
the local density passes through a maximum (purple line in Fig.~\ref{fig:o-o-dist}a). 
The duration of this high density state is about 180\,fs.  
It is worth noting that the local density is  considerably higher 
than the average within a time of $\sim$1-2\,ps  prior to and after the large-angle jump, 
revealing a density-mediated HB exchange mechanism driving 
the large-angle jumps. A jump event and the bifurcated HB state are therefore 
consequences of density fluctuations initiated some 1-2\,ps before. 
The duration 
of the bifurcated state, $\sim$180$\pm$30\,fs, coincides with the duration of the high density state 
(obtained as Full Width at Half Maximum (FWHM) of the high density peak). We also performed calculations at lower 
densities to see how the jumps change with density. Figures~\ref{fig:o-o-dist}b and ~\ref{fig:o-o-dist-gr}a show that at lower densities, 
it takes longer after the formation of the bifurcated HB state for the unstable HB pair to leave while a more stable HB partner is fully formed.   
The longer recovery times at lower densities is caused by the fact that the bifurcated state is formed 
at considerably longer distances away from the central water molecule as Fig.~\ref{fig:o-o-dist-gr}a shows. 
This  causes the newly formed O-O pair to reach equilibrium distance of 2.8~{\AA}
 (the nearest neighbour shell radius as the partial O-O pair correlation 
function in Fig.~\ref{fig:o-o-dist-gr}b shows) at longer times 
as indicated by the arrows.
Figure~\ref{fig:o-o-dist-gr}b shows the O-O pair distribution as 
a function of local density. The positions of the first and second neighbor shells are slightly 
different.

\begin{figure}
\begin{center} 
\includegraphics[viewport=100 60 760 500,width=16cm,angle=0,clip=false]
{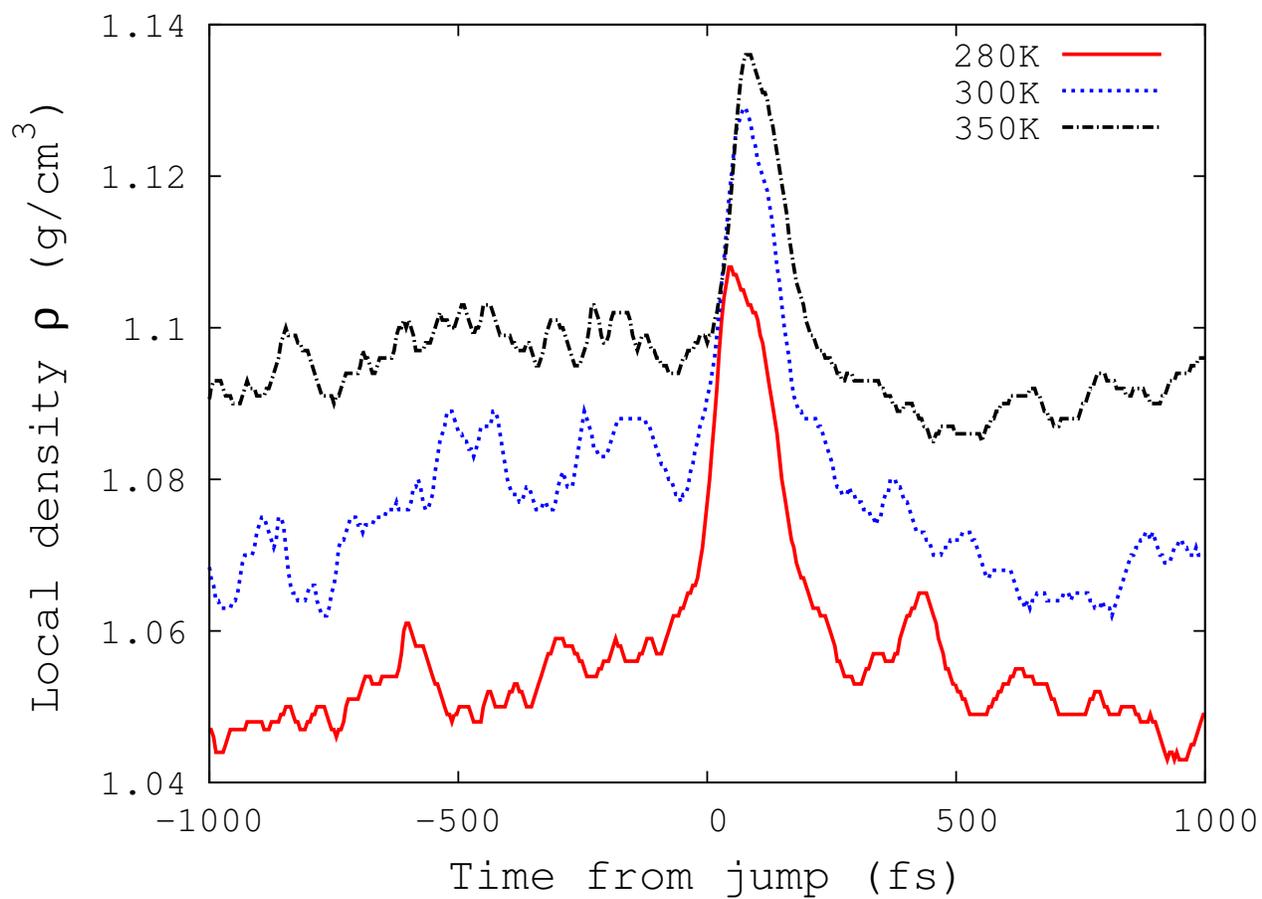}
\end{center}
\caption{Temperature effect on water's local density around the instant of large-angle jump. These are from averages over the entire simulation and all water molecules for overall average density of 1 g/cm$^3$.}
\label{fig:dens-temp-jump}
\end{figure}

Temperature plays a important role in the energy exchange process
by supplying the thermal energy necessary 
to overcome the barrier between the low and high energy HB states. 
It also causes an entropic increase that results in an increase in 
the local water density especially around the instant of the HB exchange. 
Figure~\ref{fig:dens-temp-jump}
shows the local density around the instant of an HB exchange 
at temperatures  280\,K, 300\,K and 350\,K  for bulk water. 
Although the average density of the system was fixed at 1~g/cm$^3$, 
we clearly see that as the temperature increases, the local density increases monotonically.

\section{Discussion}

We have demonstrated the  intimate relation between the hydrogen bond energy anisotropy,
local density fluctuations 
and large-angle jumps that all water molecules execute at all temperatures
in both bulk and in solvation shell around a small hydrophobic molecule (here, TMU).
By investigating the temporal properties of the frequency shift that occurs when
a water molecule switches one of its hydrogen bonds, we identified the physical origin of the
different relaxational time scales observed in different independent experiments. 
In addition, our results appear to confirm the proposed negativity track\cite{noam12}, i.e., 
the lone pairs of electrons on the oxygen atom allow for the rotational freedom of  water molecules. 
This enables the HB-donating protons to glide and enable rotation.

We have demonstrated that there is 
a strong correlation between temperature, relaxation and local density which may explain water's 
rotational slowing down in the solvation shell of apolar groups. As the local density of water increases, water molecules tend 
to rotate faster -- due to increasing number of bond bifurcations.
 This is in excellent agreement with inelastic 
ultraviolet (IUV) spectroscopy  measurements\cite{bencivenga09}.
The excluded volume
effect of the apolar group hinders overcoordinated water from forming at 
the water-hydrophobe interphase while keeping the water molecules 
(on average) tetrahedrally coordinated.

\section{Methods}

We performed \textit{ab initio} MD simulations within the Born-Oppenheimer. 
The Car-Parrinello CPMD code\cite{Car:85hr} was used with the van der Waals corrections
of Grimme\cite{grimme04,hujo2013}.
Dispersion corrections were used since it is well-known that the 
standard DFT does not contain van der Waals interactions and 
that PBE and BLYP level methods tend to produce overstructured water,
so much that such simulations are often considered to be comparable only to experiments conducted at temperatures lower 
by 50 to 100\,K\cite{czhang2011,linchun09}.
Corrections have bee shown to perform well in independent tests\cite{kruse12,linchun09,zhang2012,linchun2012,hujo2013}. 
Temperature was varied from 260 to 350\,K. For bulk water, a 54 molecule system 
was chosen. The solute-water system consisted of one TMU and 50 water 
molecules. 
The system was first equilibrated by using 
a conjugate-gradient ground state optimization of the positions.
A time-step of 5 atomic units ($\sim$0.121 fs) was used and electrons were given 
a fictitious mass of 400 au. Each of the simulations consisted of 50\,ps equilibration, 
followed by a production run of at least 100\,ps. Other details are provided in Ref.~3.

\begin{addendum}
 \item [Acknowledgements] This work has been supported by 
the Natural Sciences and Engineering Research Council of Canada (MK).
SharcNet [www.sharcnet.ca] and Compute Canada provided the computational resources. 
\end{addendum}

\newpage

\end{document}